\documentclass{article}

\usepackage{PRIMEarxiv}

\usepackage[utf8]{inputenc} 
\usepackage[T1]{fontenc}    
\usepackage{hyperref}
\usepackage{url}            
\usepackage{booktabs}       
\usepackage{bm}
\usepackage{amsfonts}       
\usepackage{amsmath}
\usepackage{amssymb}
\usepackage{nicefrac}       
\usepackage{microtype}
\usepackage{cleveref}
\usepackage{xcolor}
\usepackage{algpseudocode}
\usepackage{algorithm}
\usepackage{color,soul}
\usepackage{caption}
\usepackage{empheq}
\usepackage{fancyhdr}       
\usepackage{graphicx}       
\graphicspath{{media/}}     
\usepackage{enumitem}
\makeatletter
\newcommand{\printfnsymbol}[1]{%
  \textsuperscript{\@fnsymbol{#1}}%
}

\makeatother

\title{TOFLUX: A Differentiable Topology Optimization Framework for Multiphysics Fluidic Problems}

\author{ \and
Rahul Kumar Padhy  \\
Department of Mechanical Engineering\\
University of Wisconsin Madison\\
Madison, WI, USA \\
\texttt{rkpadhy@wisc.edu}
\and 
 Krishnan Suresh \\
 Department of Mechanical Engineering\\
 University of Wisconsin Madison\\
 Madison, WI, USA \\
\texttt{ksuresh@wisc.edu} \\
\and 
 Aaditya Chandrasekhar \\
 Department of Mechanical Engineering\\
 Northwestern University\\
 Evanston, IL, USA \\
\texttt{cs.aaditya@gmail.com} \\
}

\begin{document}
\maketitle

\begin{abstract}

Topology Optimization (TO) holds the promise of designing next-generation compact and efficient fluidic devices. However, the inherent complexity of fluid-based TO systems, characterized by multiphysics nonlinear interactions, poses substantial barriers to entry for researchers.

Beyond the inherent intricacies of forward simulation models, design optimization is further complicated by the difficulty of computing sensitivities, i.e., gradients. Manual derivation and implementation of sensitivities are often laborious and prone to errors, particularly for non-trivial objectives, constraints, and material models. An alternative solution is automatic differentiation (AD). Although AD has been previously demonstrated for simpler TO problems, extending its use to complex nonlinear multiphysics systems, specifically in fluidic optimization, is key to reducing the entry barrier.

To this end, we introduce TOFLUX, a \underline{TO} framework for \underline{flu}id devices leveraging the JA\underline{X} library for high-performance automatic differentiation. The flexibility afforded by AD enables the rapid exploration and evaluation of various objectives and constraints. We illustrate this capability through challenging examples encompassing thermo-fluidic coupling, fluid-structure interaction, and non-Newtonian flows. Additionally, we demonstrate the seamless integration of our framework with neural networks and machine learning methodologies, enabling modern approaches to scientific computing. Ultimately, the framework aims to provide a foundational resource to accelerate research and innovation in fluid-based TO. The software accompanying this educational paper can be accessed at \href{https://github.com/UW-ERSL/TOFLUX}{github.com/UW-ERSL/TOFLUX}.

\end{abstract}


\section{Introduction}
\label{sec:intro}

The simulation, design, and optimization of fluidic systems are integral to many engineering and scientific applications \cite{alexandersen2020review}, including aerospace \cite{maute2004conceptual}, biomedical systems \cite{suarez2022virtual}, HVAC \cite{manuel2018optimal}, and automotive industries \cite{kim2012topology}. These systems are governed primarily by the Navier-Stokes equations for fluid dynamics, which are often coupled with structural \cite{lundgaard2018revisiting} and thermal interactions \cite{marck2013topology}. The complexity of these multi-physics systems makes the design challenging. Thus, designers have increasingly turned to topology optimization (TO) for the design of fluidic systems \cite{alexandersen2020review}. Fundamentally, fluid TO poses the design task as an optimization problem to compute the optimal flow path given specific objectives and constraints \cite{alexandersen2020review}. TO has made the design process more automated and efficient.

Topology optimization for fluids was first introduced in the seminal paper by Borrvall and Petersson \cite{borrvall2003topology} for minimizing dissipated power in Stokes flow. Since then, fluid TO has been adapted for a variety of fluid problems, including fluid-structure interaction (FSI) \cite{lundgaard2018revisiting}, conjugate heat transfer (CHT) \cite{marck2013topology}, and non-Newtonian fluid flows \cite{pingen2010optimal, suarez2022virtual}. However, inherent to these advances is a substantial investment of time by researchers in building the necessary simulation and optimization models. Consequently, the entry barrier for new researchers into TO for fluid problems is high. In particular, fluid simulations are plagued by their nonlinearity and high computational cost \cite{fan2025diff}. As emphasized in a recent review, addressing this challenge requires educational tools that elucidate multiphysics fluid TO problems \cite{wang2021comprehensive}. In addition, fluid TO problems have the added complexity of computing sensitivities. This process involves computing the derivatives of objectives, constraints, material models, and projection or filter operations with respect to the design variables \cite{chandrasekhar2021auto}. The traditional manual sensitivity analysis process can stagnate progress, particularly given the trend toward more practical objectives and simulation models \cite{chandrasekhar2021auto}. This inevitably increases the time required for researchers to build and iterate through optimization models, making the design process inefficient.

Recently, with the emergence of machine learning, automatic differentiation (AD) has resurfaced as a powerful tool for sensitivity analysis \cite{chandrasekhar2021auto, xue2023jax}. While previous research has demonstrated to the use of AD for linear, single-physics TO problems \cite{chandrasekhar2021auto, neofytou2024level, padhy2024fluto}, its adoption and application to complex, nonlinear problems, particularly for fluid-based systems, remain scarce. Furthermore, current trends of integrating machine learning with scientific computing dictate that models be end-to-end differentiable \cite{padhy2024tomas, padhy2024photos, padhy2024voroto}.

Towards this goal, we employ JAX \cite{bradbury2018jax}, a high-performance Python library designed for end-to-end differentiability. Its NumPy-like syntax \cite{harris2020array}, low memory footprint, and support for vectorized just-in-time (JIT) compilation for accelerated code performance make it an ideal candidate for this task. In addition, we employ implicit automatic differentiation \cite{blondel2022efficient, rader2024optimistix} to handle the nonlinearities present in the system. We showcase the framework's capabilities on various fluid-based TO problems, including fluidic devices, FSI, CHT, and non-Newtonian flows. For educational purposes, we focus on simple, introductory studies in these domains. We also demonstrate integration with machine learning for design representation, showcasing the framework's utility for timely applications.

Finally, we place a great emphasis on a well-documented, consistently styled code with well-illustrated Jupyter notebooks, which are available as supplementary material to this educational paper. The readers are urged to refer to the notebooks for an involved discussion.

\section{Related Work}
\label{sec: review}

This section surveys the field of TO for fluid-based problems. The survey is structured to reflect the central contribution of this paper: the promotion of a differentiable simulation framework for designing fluidic devices. Therefore, we organize our review into two key areas. The first area covers the foundations and applications of fluid-based TO. The second area details the principles of differentiable simulation and its role in computing design sensitivities.

\subsection{Topology Optimization of Fluidic Devices}
\label{sec: review_fluid_TO}

Fluidic devices play a central role in systems involving material and energy transport. The optimization of such transport systems is critical in established fields such as automotive \cite{dede2014TOCar}, HVAC \cite{sun2025HXTO}, and aerospace \cite{maute2004AeroTO}, as well as in emerging areas such as soft robotics \cite{dalklint2024InflatableRobotTO} and battery technology \cite{kim2012TOFuelCell}. Given their complex and multiphysics nature, TO is emerging as a method for the automated design of these fluidic systems. This review provides an overview of seminal work in fluid-based TO, with a particular emphasis on non-Newtonian fluids, conjugate heat transfer, and fluid-structure interaction. For a more comprehensive review of the field, we refer readers to Alexandersen et al. \cite{alexandersen2020review}.

The foundation of TO for fluid flow was laid by Borrvall and Petersson, who formulated a density-based approach to minimize pressure drop \cite{borrvall2003topology}. Their approach utilized the Stokes equation with a Brinkman–Darcy penalization for low-Reynolds number conditions. This methodology was later extended by Gersborg-Hansen et al. \cite{gersborg2005topology} to applications in microfluidics and micro-electro-mechanical systems. Subsequent research has focused on refining this approach. For instance, Guest et al. \cite{guest2006topology} proposed a Darcy–Stokes finite element model in lieu of the Brinkman penalization to produce distinct void-solid topologies without artificial material regions.

Furthermore, Haubner et al. \cite{haubner2023novel} promoted the use of fractional-order Sobolev spaces to improve numerical stability and convergence. Another refinement involved geometric representation, where Jensen et al. proposed anisotropic mesh adaptation for higher-fidelity descriptions \cite{jensen2018topology}. Alongside these numerical improvements, the physical modeling itself has also evolved. For instance, Wiker et al. \cite{wiker2007topology} treated viscosity as a design-dependent parameter, and Shen et al. \cite{shen2018topology} developed three-phase interpolation models.

The adaptation of these various methods has been significantly aided by the availability of educational and open-source codes \cite{wang2021comprehensive}. One such code, the MATLAB-based PolyTopFluid, presented a stable, low-order discretization of the Stokes equations using polygonal finite elements \cite{pereira2016fluid}. Another comprehensive educational resource was provided by Alexandersen \cite{alexandersen2023detailed} for the density-based TO of Navier-Stokes problems. A distinctive characteristic of fluidic problems is their high computational cost. To address this, Liu et al. presented a framework for large-scale optimization of fluidic devices on GPUs \cite{liu2022marker}. Other educational resources that advance the field include the MATLAB code from Kumar et al. \cite{kumar2023topress} for optimizing structures with design-dependent pressure loads.

Building on these single-physics foundations, the field has expanded to tackle coupled multiphysics systems. Addressing such systems is crucial for solving complex, real-world engineering challenges. For instance, Maute et al. \cite{maute2004conceptual} performed a dry fluid-structure interaction (FSI) optimization of an interior wing design, where the fluid-solid interface was treated as constant. In contrast, wet TO-FSI was pioneered by Yoon et al. \cite{yoon2010topology}, who used density-based TO to minimize mechanical compliance and pressure drop. A recent FSI method implicitly defines the boundary using the density jump between elements, adding a pressure penalty to improve physical accuracy \cite{hederberg2025fluid}.

Another primary multiphysics application is the design of high-performance thermal management systems. These systems often require conjugate heat transfer (CHT) models to optimize the trade-off between maximizing thermal performance and minimizing pressure drop \cite{marck2013topology, subramaniam2019topology}. This CHT-based TO approach has been used to design components such as heat exchangers \cite{matsumori2013topology} and microchannel heat sinks \cite{koga2013development}. Finally, designing systems for non-Newtonian fluids, which are prevalent in bioengineering \cite{suarez2022virtual}, is of emerging interest. Modeling non-Newtonian fluids requires coupling flow equations with appropriate rheological models, which often involves application-specific objectives such as minimizing hemolysis in blood-flow devices \cite{alonso2021topology}.

\subsection{Differentiable Simulation for Sensitivity Analysis}
\label{sec: review_differentiableSim}

In TO, the algorithms commonly employ gradient-based optimizers such as the Method of Moving Asymptotes (MMA) \cite{svanberg1987method}. These optimizers utilize the sensitivities or gradients of the objective and constraints with respect to the design variables. There are conceptually four methods for computing these sensitivities \cite{baydin2018automatic}: (1) numerical, (2) symbolic, (3) manual, and (4) automatic differentiation (AD). The numerical method, based on finite differences, suffers from truncation and floating-point errors and is therefore not recommended. The symbolic method, using software such as SymPy, is impractical for code containing loops or conditional logic.

The manual method is often the default approach for obtaining sensitivities. However, the manual process of deriving sensitivities is cumbersome, error-prone, and a bottleneck in research and development \cite{chandrasekhar2021auto}. To overcome these limitations, AD is increasingly being used \cite{chandrasekhar2021auto}. This paper promotes the use of AD for computing sensitivities in fluidic TO problems. AD comprises a set of methods that efficiently and precisely calculate derivatives of functions implemented as computer programs. Although while AD is relatively new to topology optimization, it has been established for decades \cite{rumelhart1986learning} and has seen diverse applications from molecular dynamics to photonics \cite{schoenholz2019jax, minkov2020inverse, padhy2024fluto, chen2025uncertainty}.

The application of AD is well-established in finite element analysis \cite{ozaki1995higher}. Recent studies have extended its use to shape optimization tasks within frameworks such as Firedrake \cite{paganini2021fireshape} and to TO problems involving turbulent flow \cite{dilgen2018topology}. For transient flow, specific AD tools such as CoDiPack \cite{sagebaum2019high} and Tapenade \cite{hascoet2013tapenade} have proven effective \cite{norgaard2017applications}. Beyond standalone tools, comprehensive platforms such as OpenMDAO integrate AD by automating total derivative calculations \cite{gray2019openmdao}. The automation provided by AD has also been instrumental in integrating machine learning (ML) into TO workflows. Specifically, the combination of AD with differentiable ML platforms such as PyTorch \cite{paszke2019pytorch} and JAX \cite{bradbury2018jax} allows neural networks to serve as surrogate models \cite{padhy2024voroto}, design representations \cite{chandrasekhar2021tounn}, or generative latent spaces \cite{padhy2024tomas}.

\section{Fluid Flow Optimization}
\label{sec:flow}

In this section, we consider the design of fluidic devices with the objective of minimizing the dissipated power. We assume that the flow is governed by the steady state incompressible Navier-Stokes equations. Furthermore, we impose a constraint on the lower bound on the solid/material volume. We refer the readers to the work of \cite{alexandersen2023detailed} for a detailed treatment of this subject.

The fluid is governed  by \cref{eq:momentum,eq:continuity}, which incorporates a design dependent Brinkman penalty term $(\alpha (\gamma) \bm{u})$ \cite{alexandersen2023detailed} to facilitate topology optimization. This term penalizes fluid flow in solid regions within the design domain.

\begin{subequations}
\begin{empheq}[left={R(\gamma, \mathbf{u}, p) = \empheqlbrace}]{align}
    \varrho (\mathbf{u} \cdot \nabla) \mathbf{u} - \nabla \cdot \left( \mu (\nabla \mathbf{u} + \nabla \mathbf{u}^T) \right) + \nabla p - \mathbf{\alpha} (\gamma) \mathbf{u} &= \mathbf{0}, \label{eq:momentum} \\
    \nabla \cdot \mathbf{u} &= 0, \label{eq:continuity}
\end{empheq}
\end{subequations}

Here, $\gamma \in [0,1]$ denotes the pseudo-density design variable, with $\gamma=0$ and $\gamma=1$ representing fluid and material respectively. Furthermore,  $\bm{u}$ is the velocity field, $p$ is the pressure, $\varrho$ is the fluid density, $\mu$ is the dynamic viscosity and $\alpha$ is the design-dependent Brinkman penalty. Additionally, to map the design variable/pseudo-densities to the Brinkman penalty term, we use a RAMP interpolation \cref{eq:ramp_penalty}, controlled by the penalization parameter $q_{\alpha}$, following the work of \cite{alexandersen2023detailed}. Observe that the Navier-Stokes equations are nonlinear owing to the convective term $\varrho (\mathbf{u} \cdot \nabla) \mathbf{u}$. This results in nonlinear finite element equations ($\mathbf R(\bm{\gamma}, \mathbf u, p)$) that is solved iteratively using the Newton-Raphson algorithm.

\begin{equation}
    \alpha(\gamma) = \alpha_{\min} + (\alpha_{\max} - \alpha_{\min}) \frac{\gamma}{1 + q_{\alpha} - q_{\alpha}\gamma} \label{eq:ramp_penalty}
\end{equation}

Furthermore, we aim to minimize the dissipated power, subject to a constraint on the material volume fraction. The design variables, $\gamma_e$, are the pseudo-densities of the elements in a mesh with Q1-Q1 elements. The optimization problem is formulated as follows: 

\begin{subequations}
\begin{align}
\min_{\bm{\gamma}}\;\; & J
      = \int_{\Omega}
        \tfrac12\,(\mu\,\bigl(\nabla\mathbf u
        +\nabla\mathbf u^{\mathsf T}\bigr)\!:\!
        \bigl(\nabla\mathbf u+\nabla\mathbf u^{\mathsf T}\bigr) \ + \alpha (\gamma)\,\mathbf u\!\cdot\!\mathbf u )\;d\Omega  \label{eq:obj_dissip_pow} \\
\text{s.\,t.}\;\;
& \mathbf R(\bm{\gamma}, \mathbf u, p) = \mathbf 0 \label{eq:dissip_pow_pde}  \\
& g(\bm{\gamma})=1-\frac{\displaystyle
             \sum_e \gamma_e\,v_e}{V^{\ast}} \le 0 \label{eq:dissip_pow_vol_cons} 
\end{align}
\label{eq:dissip_power_opt}
\end{subequations}

\begin{figure}[H]
 	\begin{center}
		\includegraphics[scale=0.57,trim={0 0 0 0},clip]{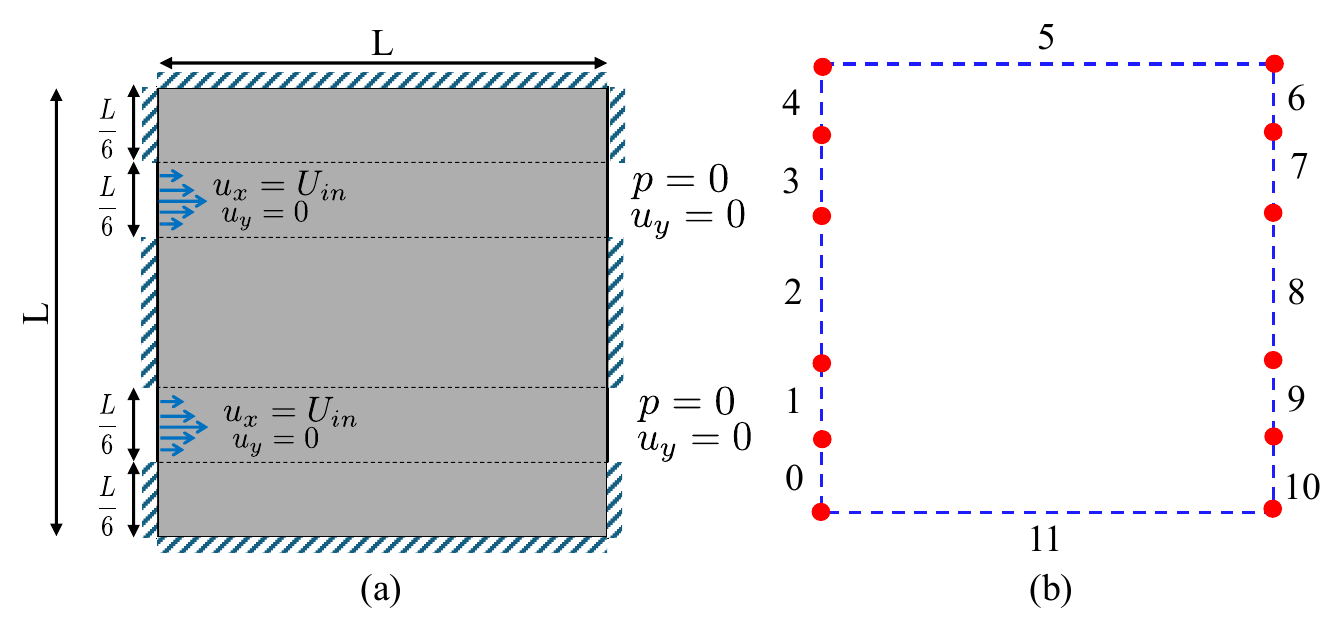}
 		\caption{(a) Double pipe design domain $(L=1)$ and boundary conditions. (b) Boundary segments of the geometry}.
        \label{fig:double_pipe_bc}
	\end{center}
 \end{figure}

\begin{figure}[H]
 	\begin{center}
		\includegraphics[scale=0.57,trim={0 0 0 0},clip]{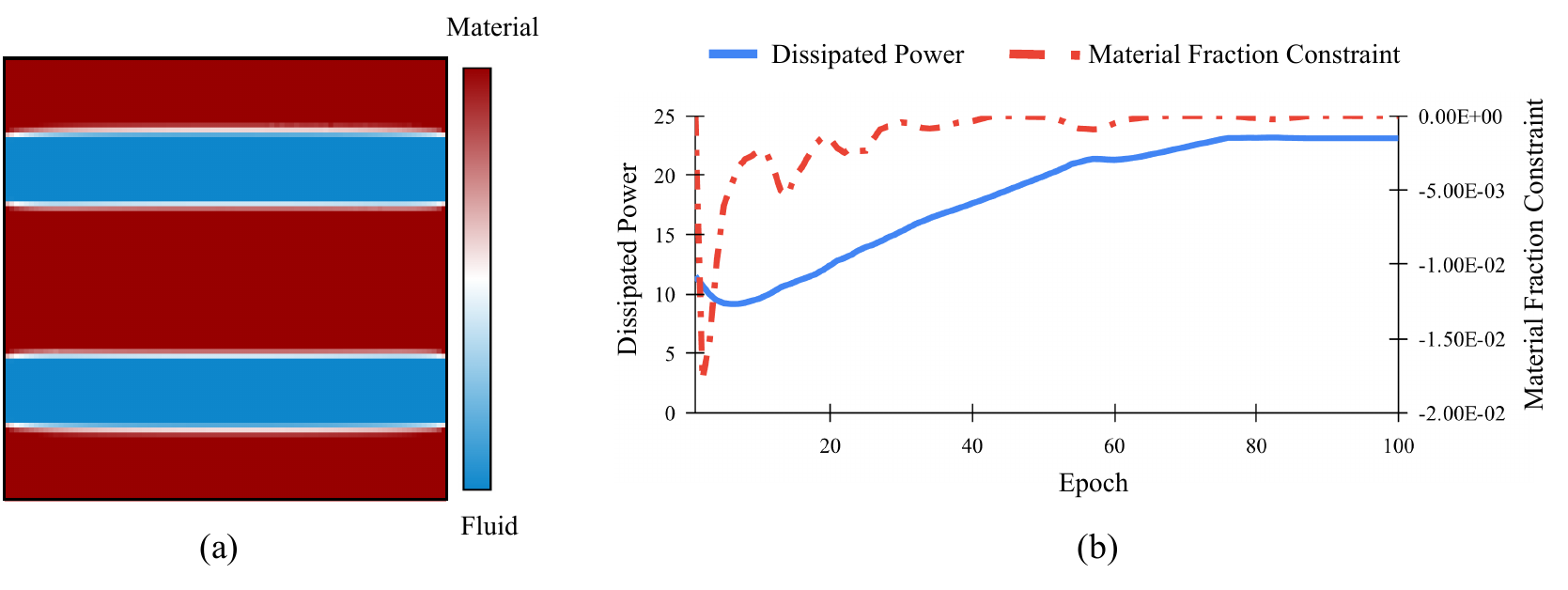}
 		\caption{(a) Optimized design with dissipated power of $22.1$. (b) Convergence of dissipated power and material fraction constraint.}.
        \label{fig:double_pipe}
	\end{center}
 \end{figure}

As an example, consider minimizing the dissipated power of a flow through a double pipe; see \Cref{fig:double_pipe_bc}a. We once again refer the readers to the exhaustively documented Jupyter notebooks for a detailed description. Additionally, the repository's \texttt{README} provides a concise index of all notebooks, indicating their locations and explaining their function. Here, we summarize the key steps captured in the Jupyter notebook (that the reader is encouraged to follow):

\begin{enumerate}[itemsep=2pt, parsep=2pt]
    \item \textbf{Geometry}: As the first step, one must create a geometry where various segments are identified as in \Cref{fig:double_pipe_bc}b. In \texttt{TOFLUX}, the geometry is captured in a simple boundary representation form via a \texttt{JSON} file where the boundary segments are numbered. These segments will be used later for specifying boundary conditions.
    
    \item \textbf{Meshing}: Given the geometry, we create a (voxel grid) mesh, by specifying the number of elements desired along x and along y. Here we use a $ 90 \times 90$ mesh.
    
    \item \textbf{Material}: Next the appropriate materials are assigned. Here, we assign a mass density $(\varrho)$ of 1 and a dynamic viscosity $(\mu)$ of 1.
    
    \item \textbf{Boundary Conditions}: Boundary conditions are specified using the segment numbers identified in Figure \ref{fig:double_pipe_bc}b. For example, parabolic inlet velocities are prescribed on segments 1 and 3, while outlet pressure conditions are prescribed on segments 7 and 9, and so on.
    
    \item \textbf{Solvers}: The Newton-Raphson method (employed for solving the non-linear fluid flow problem) relies on a linear solver. We integrate our JAX framework with various linear solvers, including PETSc \cite{balay2019petsc}, pyamg \cite{bell2022pyamg}, and PARDISO \cite{schenk2001pardiso}.
    
    \item \textbf{Objective and Gradient}: The objective (to minimize) is the dissipated power. The gradient of the objective function (with respect to the pseudo-density variables) is obtained via JAX's \texttt{value\_and\_grad}, i.e., the user need not supply the gradient expression. The JAX environment automatically traces the chain of function calls and ensures end-to-end automatic differentiation. Appendix (\cref{sec:appendix_implicitAutoDiff}) describes how the gradients are computed for nonlinear problems using implicit automatic differentiation. 
    
    \item \textbf{Constraint and Gradient}: Similarly, one defines the constraint; in this case, a simple volume constraint of $V^* = 0.67$. Once again, we rely on JAX to compute the gradient.
    
    \item \textbf{Optimize}: Here, we rely on MMA \cite{svanberg1987method} for optimization, where we supply the objective, constraint and their gradients. The final topology is illustrated in \Cref{fig:double_pipe} (a), matching those that obtained in the literature \cite{alexandersen2023detailed}. Additionally, the convergence is illustrated in \cref{fig:double_pipe}(b).
    
\end{enumerate}

\subsection{Flow Reversal}
\label{sec:flow_flowRev}

We now explore some of the advantages of the AD framework.  Traditionally, researchers spend significant time not only formulating the objective and constraints but also in deriving and implementing the corresponding sensitivities. The researcher's task is simplified when adopting a differentiable programming framework; allowing them to focus only on the forward models.

\begin{figure}[H]
 	\begin{center}
		\includegraphics[scale=0.57,trim={0 0 0 0},clip]{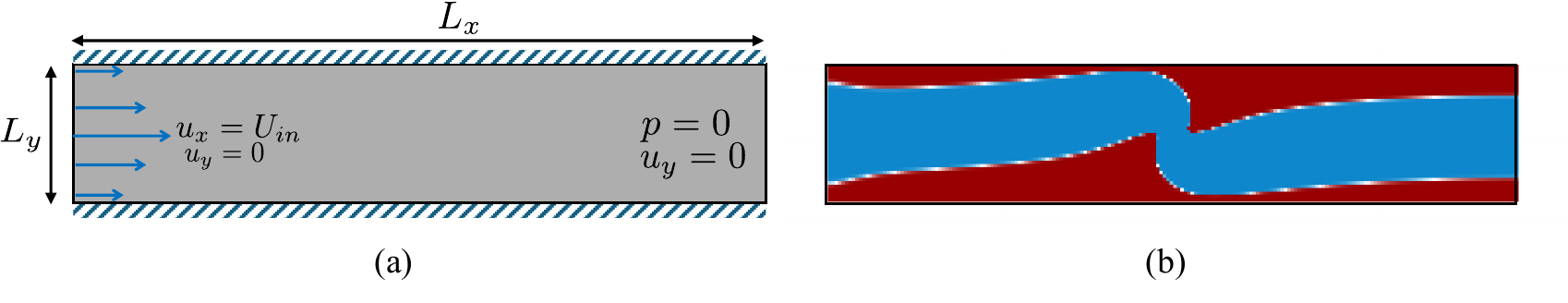}
 		\caption{(a) Flow reversal design domain $(L_x=5, L_y =1)$ and boundary conditions. (b) Optimized vertical velocity at center = $2.47$}
        \label{fig:flow_reversal}
	\end{center}
 \end{figure}

For instance, consider an extension to the above fluid optimization problem where we now seek to reverse the flow with a constraint on the maximum dissipated power; see \cref{fig:flow_reversal}(a). One can then reformulate the problem in \Cref{eq:dissip_power_opt} with an objective to maximize the negative of vertical velocity at domain's center($L_x/2, L_y/2$) along with material fraction and dissipated power constraints (\Cref{eq:flow_reverse_opt}).

\begin{subequations}
\begin{align}
\max_{\bm{\gamma}}\;\; &  -u_{\text{y}}(L_x/2, L_y/2)  \label{eq:obj_flow_rev} \\
\text{s.\,t.}\;\;
& \mathbf R(\bm{\gamma}, \mathbf u, p) = \mathbf 0 \label{eq:flow_rev_pde}  \\
& g_v(\bm{\gamma}):= 1-\frac{\displaystyle
             \sum_e \gamma_e\,v_e}{V^{\ast}} \le 0 \label{eq:flow_rev_vol_cons} \\
& g_d(\bm{\gamma}, \bm{u}) := \Phi(\bm{\gamma}, \bm{u}) - C_f \Phi_0 \leq 0 \label{eq:flow_rev_dissip_pow_cons} 
\end{align}
\label{eq:flow_reverse_opt}
\end{subequations}

Here, $\Phi$ is the dissipated power of the current design, $\Phi_0$ is the dissipation in an empty channel, volume constraint of $V^* = 0.4$ and $C_f (=5)$ is a user-defined factor. Once again, we refer the readers to \cite{alexandersen2023detailed,gersborg2005topology} and our code for a detailed discussion. The optimized design (\Cref{fig:flow_reversal}) and performance are obtained and found to be similar to those reported in the the above mentioned references \cite{alexandersen2023detailed,gersborg2005topology}. We also observe that the optimized channel induces a vertically downward velocity at the domain center.

\subsection{Drag Minimization}
\label{sec:flow_dragMin}

\begin{figure}[H]
 	\begin{center}
		\includegraphics[scale=0.5,trim={0 0 0 0},clip]{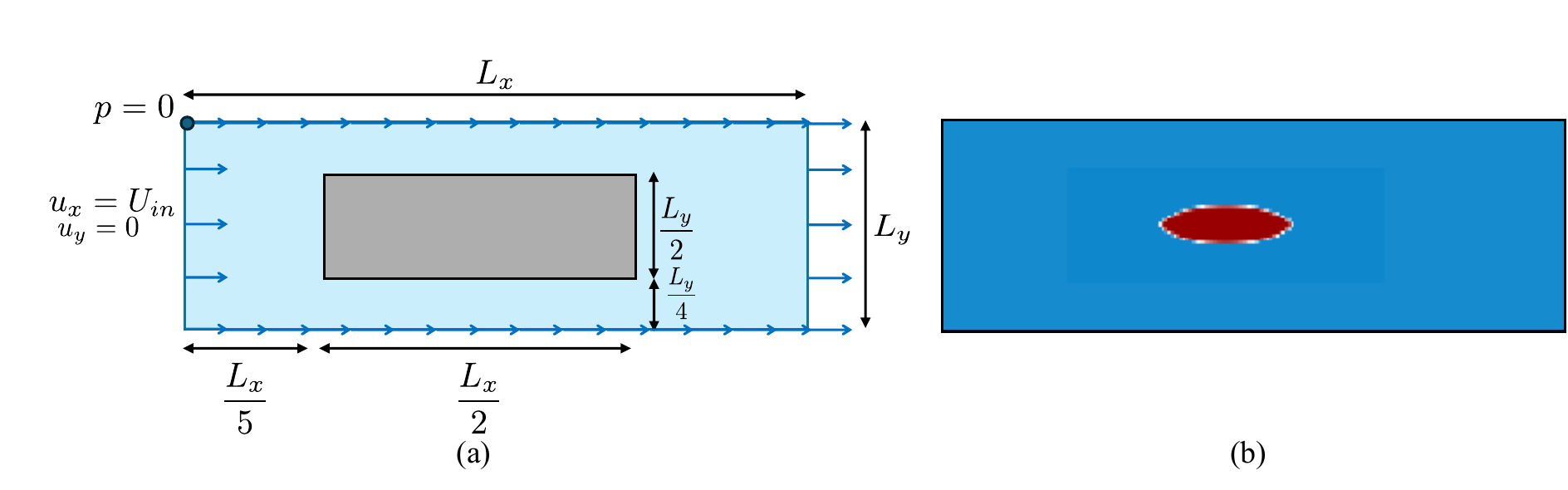}
 		\caption{(a) Drag minimization design domain $(L_x=3, L_y =1)$ and boundary conditions. The region in gray is the design region. (b) Optimized design with minimum drag.}
        \label{fig:drag}
	\end{center}
 \end{figure}

Another common application of fluid-based TO is the design of optimal aerodynamic structures. This task is typically accomplished by formulating objectives and constraints based on aerodynamic forces such as drag and lift. As a representative example, we consider the minimization of drag on an object subject to a material volume constraint shown in \cref{eq:drag_opt}. The drag is computed by integrating the body force over the flow domain (\cref{eq:drag_int}). Here, $\alpha$ is the design dependent Brinkmann penalty, $u_x$ is the horizontal velocity in the domain and volume constraint of $V^* = 0.15$ w.r.t the whole domain. Then, optimizing subject to boundary condition as illustrated in \Cref{fig:drag}(a), we obtain an optimal aerodynamic design (\Cref{fig:drag}(b)) that agrees with the result in \cite{alexandersen2023detailed}.

\begin{subequations}
\begin{align}
   \min_{\bm{\gamma}}\;\; & \text{Drag} = \int_{\Omega} \alpha\,u_{x}\, d\Omega \label{eq:drag_int} \\
    & \mathbf R(\bm{\gamma}, \mathbf u, p) = \mathbf 0 \label{eq:drag_pde}  \\
& g(\bm{\gamma})=1-\frac{\displaystyle
             \sum_e \gamma_e\,v_e}{V^{\ast}} \le 0 \label{eq:drag_vol_cons} 
\end{align}
\label{eq:drag_opt}
\end{subequations}

\subsection{Neural Representation}
\label{sec:flow_tounn}

A powerful paradigm in research today is scientific machine learning, where one combines traditional scientific computing with modern machine learning (ML) tools. In addition to supporting AD, JAX is a powerful ML platform. By adopting the same framework for our simulation and design optimization, we can leverage ML directly and easily in our TO framework.

\begin{figure}[H]
 	\begin{center}
		\includegraphics[scale=0.5,trim={0 0 0 0},clip]{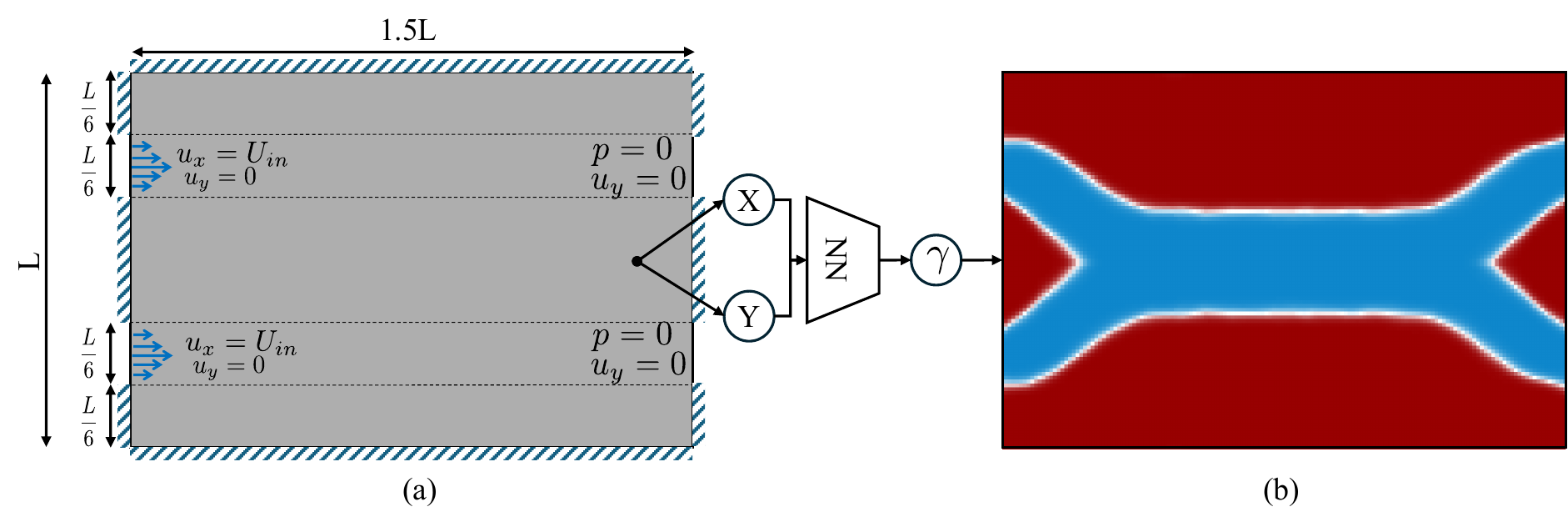}
 		\caption{(a) Double pipe design domain $(L=1)$ and boundary conditions. (b) Optimized design with dissipated power of $22.1$}.
        \label{fig:tounn}
	\end{center}
 \end{figure}

For instance, we consider integrating the neural reparameterization framework as proposed by \cite{chandrasekhar2021tounn}. Here, we use the weights of a neural network, instead of element densities, to parameterize our design. This approach allows for a more expressive and continuous design space. We once again consider the double pipe experiment, except that the domain length is $1.5L$, as shown in \Cref{fig:tounn}(a), with the objective of minimizing dissipated power subject to a volume constraint.

The loss is formulated using a log-barrier scheme to combine the objective (dissipated power) and constraint (material fraction), which is then minimized using the ADAM optimizer \cite{kingma2014adam}. We refer the readers to \cite{chandrasekhar2021tounn} for a detailed discussion on this neural network formulation. Critically, we require the solver to also be in the JAX framework and differentiable to obtain the gradients of the loss with respect to the network parameters. Furthermore, the neural network and optimizers such as ADAM are readily available within the JAX environment; streamlining the workflow. The resulting design is shown in \Cref{fig:tounn}(b). This experiment showcases that our framework is readily adaptable for scientific machine learning research, opening up avenues for more complex and data-driven design problems.

\section{Non-Newtonian Fluids}
\label{sec:nonNewton}

While the governing Navier-Stokes equations are themselves nonlinear, many problems feature additional nonlinearities such as from the fluid rheological model. A primary example is the behavior of non-Newtonian fluids, which are common in applications such as fluidic devices, biomedical systems (e.g., blood flow), additive manufacturing (e.g., polymers), and industrial mixers (e.g., emulsions). For these materials, the viscosity is often modeled as a function of the shear rate. We exhibit the ease of integrating such nonlinear behavior into our framework. We simply have to define the forward model incorporating the effect of the shear rate on viscosity; once again, the sensitivities are derived automatically. 

\begin{figure}[H]
 	\begin{center}
		\includegraphics[scale=0.57,trim={0 0 0 0},clip]{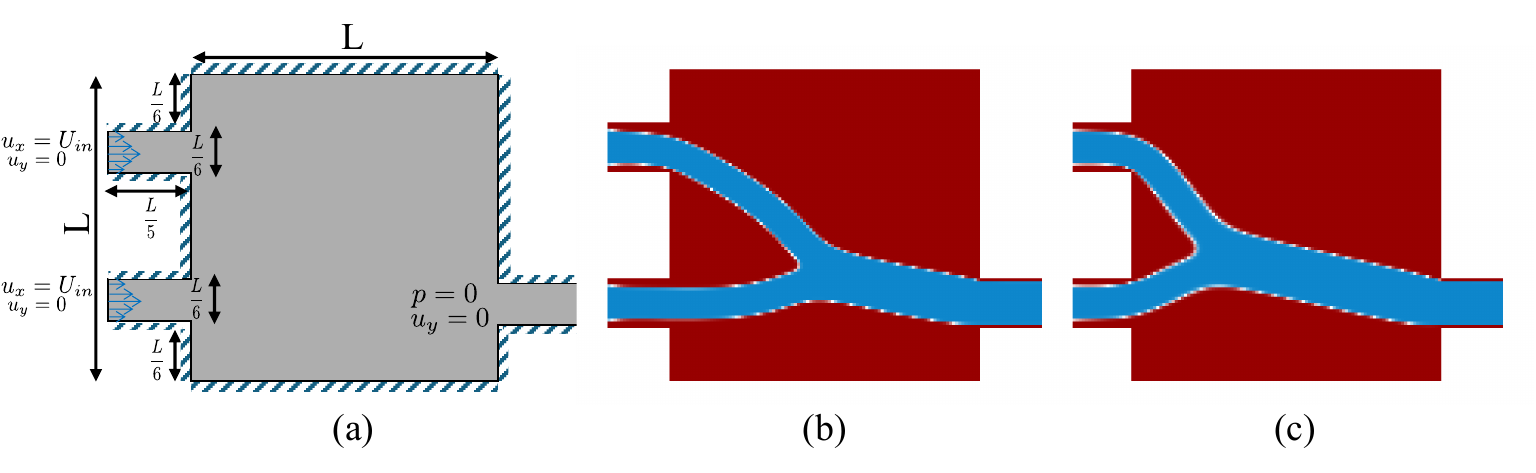}
 		\caption{(a) Bifurcated artery design domain $(L=1)$ and boundary conditions.  Optimized design with (dissipated power and volume constraint of $0.76$) for (b) Newtonian fluid $(6.47)$ and (c) Non Newtonian fluid $(10.54)$.}
        \label{fig:non_newton}
	\end{center}
 \end{figure}

To demonstrate, we use the arterial bypass example from Suarez et al \cite{suarez2022virtual}. The domain and boundary conditions with fixed and graft arteries are illustrated in \cref{fig:non_newton} (a). We solve the problem for two  different fluid models:

\begin{enumerate}[itemsep=2pt, parsep=2pt]
\item \textbf{Newtonian Fluid:} We first solve the problem using a simple Newtonian fluid with a viscosity of $3.45\times 10^{-2}$ Pa·s. The resulting design is illustrated in \cref{fig:non_newton}(b).

\item \textbf{Carreau-Yasuda Model:} We then change the material model to the Carreau-Yasuda model (\cref{eq:carreau-yasuda}, which is well-suited for blood rheology. The parameters used are $\mu_{\infty}=3.45\times 10^{-2} $ Pa.s, $\mu_{0}=0.8$, $a=2.0$, $\lambda = 3.31$ and $n=0.3568$. The resulting design is shown in \cref{fig:non_newton}(c). We observe the host-artery–graft junction shifts closer to the inlet in the non-Newtonian case due to the higher flow resistance.

\begin{equation}
\mu(\dot{\varepsilon})
= \mu_{\infty}
+ \bigl(\mu_{0}-\mu_{\infty}\bigr)
\left[\,1+(\lambda\,\dot{\varepsilon})^{a}\right]^{\tfrac{n-1}{a}} ,
\label{eq:carreau-yasuda}
\end{equation}
\end{enumerate}

These experiments illustrates our framework's ability to handle complex material nonlinearities with ease; enabling researchers to experiment and formulate design optimization for more complex problems rapidly. Finally, we also highlight that there is interest in exploring ML surrogate models for modeling fluid rheological behavior, which, once again, can be readily integrated into our framework \cite{reyes2021learning}.

\section{Fluid Structure Interaction}
\label{sec:fsi}

Next, we extend our discussion to multiphysics problems. In particular, we consider a weakly coupled Fluid-Structure Interaction (FSI) problem, where a flowing fluid imparts forces on a solid structure. Our differentiable framework enables straightforward computation of the necessary adjoints and gradients. We refer the reader to \cite{lundgaard2018revisiting} for a detailed treatment of this topic.

The structural response is governed by the Navier-Cauchy equations for linear elasticity. These equations are given as:

\begin{subequations}
\begin{align}
\nabla \cdot \bm{\sigma}^s + \mathbf{f} &= \mathbf{0} && \text{in } \Omega, \label{eq:cauchy_momentum} \\
\bm{\sigma}^s &= \mathbf{C} : \bm{\epsilon}^s, \label{eq:cauchy_constitutive} \\
\bm{\epsilon}^s &= \frac{1}{2} \left( \nabla \mathbf{d} + (\nabla \mathbf{d})^{\mathsf{T}} \right), \label{eq:cauchy_strain} \\
\bm{\sigma}^s \cdot \mathbf{n} &= -p && \text{on fluid-solid interface}. \label{eq:cauchy_pressure_bc}
\end{align}
\end{subequations}

where, $\bm{\sigma}^s$ is the Cauchy stress tensor, $\mathbf{f}$ is the body‑force vector. The structural displacement field is denoted by $\mathbf{d}$ and $\bm{\epsilon}^s$ is the infinitesimal strain tensor. $\mathbf{C}$ is the fourth-order elasticity tensor. The fluid pressure $p$ is applied as a Neumann boundary condition, where $\mathbf{n}$ is the outward-pointing normal vector. In our optimization problem, the elasticity tensor and the pressure are design-dependent terms.

The weakly coupled FSI problem is solved using the following sequential steps:
\begin{enumerate}[itemsep=2pt, parsep=2pt]
\item First, the fluid flow field is computed using the Navier-Stokes equations detailed in \cref{eq:momentum} and \cref{eq:continuity}.
\item Next, the pressure field from the fluid solution is used to compute a coupling force on the fluid-solid interface.
\item Finally, this force is used as an input to solve for the structural displacement field using the Navier-Cauchy equations.
\end{enumerate}

\begin{figure}[H]
 	\begin{center}
		\includegraphics[scale=0.57,trim={0 0 0 0},clip]{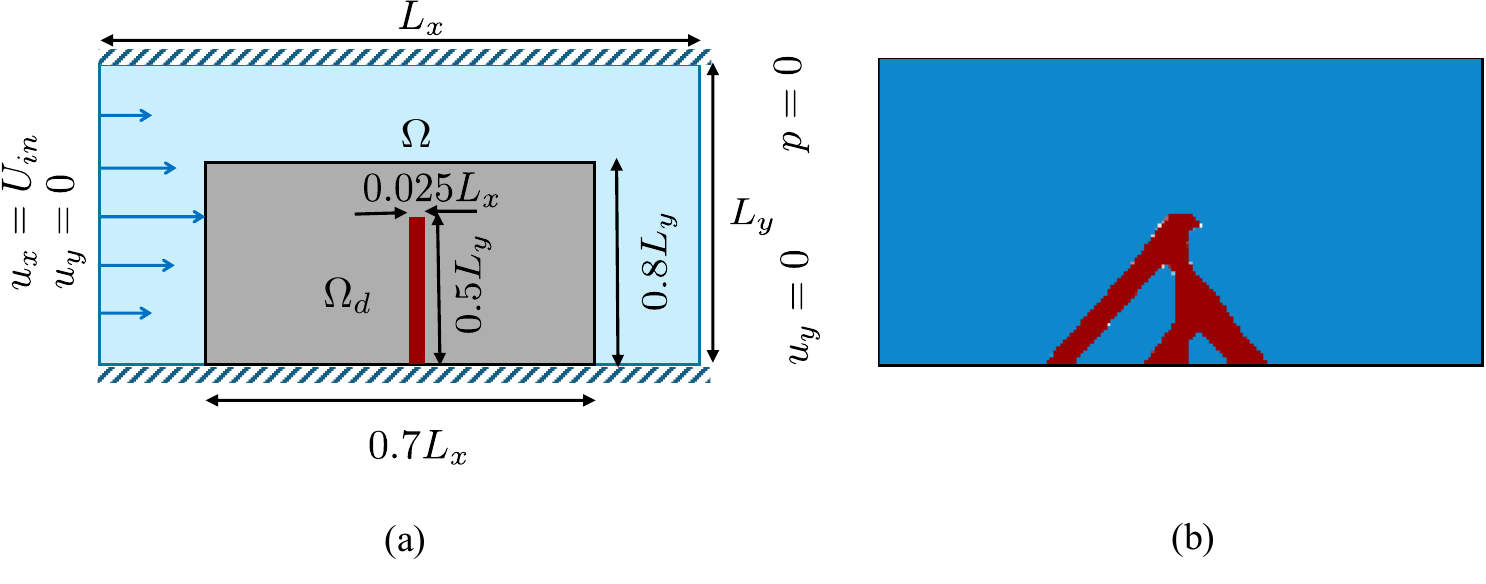}
 		\caption{(a) Channel boundary conditions with wall design  domain $(L_x=2)$ and $(L_y=1)$. (b) Optimized design for minimum compliance.}
        \label{fig:fsi}
	\end{center}
 \end{figure}

The above steps are repeated within the framework to find the optimal structure. We demonstrate our framework by minimizing the structural compliance of a wall subjected to fluid flow and volume fraction of $0.1$ w.r.t the design domain as shown in \cref{eq:fsi_opt}. Here $\mathbf{R}$ is coupled FSI residual and $V^*$ is the maximum volume fraction. The problem domain and boundary conditions are shown in \cref{fig:fsi}(a). We consider a fluid with a Reynolds number of 1, a mass density of 1, and a dynamic viscosity of 1. The solid wall and any added material are modeled as a linear elastic solid with a Young’s modulus of  $E = 10^5$ and a Poisson’s ratio of 0.3. During the optimization, material is added around the wall, which has a constant density, to maximize stiffness. The optimized design, shown in \cref{fig:fsi}(b), aligns with results from existing literature. Although drag is not included as a stand-alone objective, the coupled multiphysics design is driven by two criteria: minimizing structural drag and maximizing structural stiffness. The results demonstrate the framework's capability to handle coupled multiphysics problems effectively.

\begin{subequations}
\begin{align}
   \min_{\bm{\gamma}}\;\; & \text{Compliance} = \mathbf{f^Td},\label{eq:fsi_obj} \\
    & \mathbf R(\bm{\gamma}, \mathbf u, p, \mathbf d) = \mathbf 0 \label{eq:fsi_pde}  \\
& g(\bm{\gamma})=\frac{\displaystyle
             \sum_e \gamma_e\,v_e}{V^{\ast}} - 1\le 0 \label{eq:fsi_vol_cons} 
\end{align}
\label{eq:fsi_opt}
\end{subequations}

\section{Thermo-Fluid}
\label{sec:thermoFluid}

Finally, we apply the framework to thermo-fluid devices involving conjugate heat transfer (CHT). CHT couples heat conduction in solids with convective heat transport in fluids. This requires solving the steady incompressible Navier-Stokes equations alongside a thermal energy transport equation.

We pose the optimization as a multi-objective formulation balancing two conflicting goals: minimizing fluid power dissipation and maximizing thermal power recovery. Each objective is first normalized using reference values from single-objective optima to ensure a comparable scale \cite{marck2013topology, subramaniam2019topology}.

\begin{figure}[H]
 	\begin{center}
		\includegraphics[scale=0.4,trim={0 0 0 0},clip]{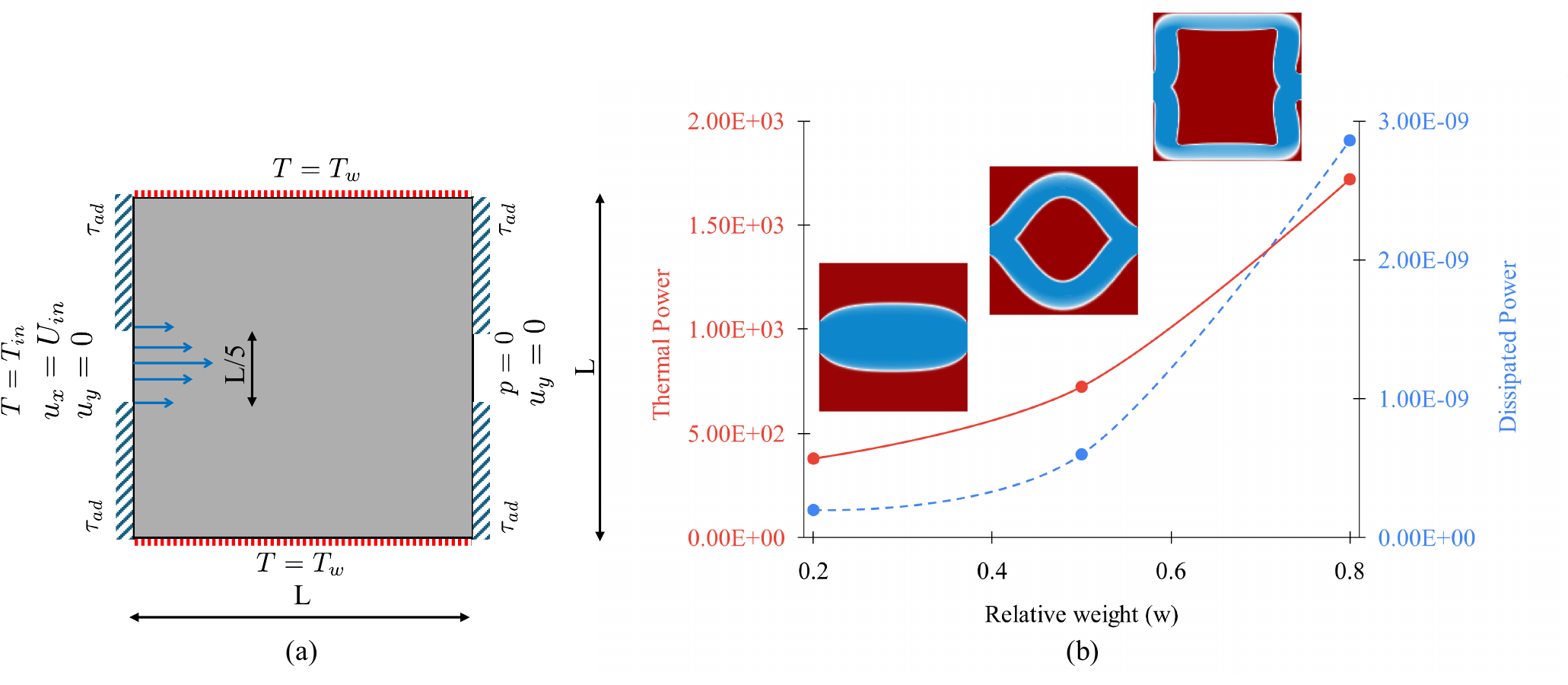}
 		\caption{(a) Conjugate heat-transfer design domain and optimal designs with different objective weights $w$: (b) $w = 0.2$, (c) $w = 0.5$ and (d) $w = 0.8$}.
        \label{fig:cht}
	\end{center}
 \end{figure}

We illustrate the optimziation with a benchmark CHT problem in a 2D domain as shown in (\Cref{fig:cht}(a)). The hot walls ($T_w = 10 ^{\circ} C$) heat a cold fluid (with specific heat capacity of $C_p$) flowing from a $0^{\circ} C$ inlet. This setup creates a multiphysics scenario where heat conducts through solids and is advected by the fluid. The analysis solves the coupled steady-state incompressible Navier-Stokes with (Re=3) (\Cref{eq:momentum,eq:continuity}) and the energy equation (\Cref{eq:cht_governing}). 

\begin{equation}
    \mathbf{u} \cdot \nabla T - \nabla \cdot \left(\frac{1}{\text{Pe}(\gamma)}\nabla T\right) = 0 \quad \text{in } \Omega \label{eq:cht_governing}
\end{equation}
Here Pe is the design dependent P\'eclet number and T is temperature.
The fluid objective ($J_f$) is the dissipated power (\Cref{eq:obj_dissip_pow}), while the thermal objective ($J_t$) is the net convective heat flow between the outlet and inlet (\Cref{eq:recoverable_power}). Here $\mathbf{n}$ denotes the outward unit normal on the boundary. The combined objective optimization is formulated as in \Cref{eq:CHT_opt}.

\begin{equation}
    J_t(\mathbf{u}, T) = \int_{\Gamma} \mathbf{n} \cdot \mathbf{u} (\varrho C_p T) d\Gamma
    \label{eq:recoverable_power}
\end{equation}

\begin{subequations}
\begin{align}
   \min_{\bm{\gamma}}\;\; &  (1-w)J_f - w J_t,\label{eq:CHT_obj} \\
    & \mathbf R(\bm{\gamma}, \mathbf u, p, T) = \mathbf 0 \label{eq:fsi_pde}  \\
& g(\bm{\gamma})=1 -\frac{\displaystyle
             \sum_e \gamma_e\,v_e}{V^{\ast}} \le 0 \label{eq:CHT_vol_cons} 
\end{align}
\label{eq:CHT_opt}
\end{subequations}
Here, $\mathbf{R}$ is the coupled CHT residual and $V^{\ast} = 0.4$ is the volume constraint.
Adjusting the weight $w$ generates a Pareto front of optimal designs \Cref{fig:cht}(b), from fluid-optimal to thermally-optimal. We observe that for $w=0.2$—where dissipated power carries greater weight—the optimizer yields a nearly straight channel. As $w$ increases (placing more weight on recoverable thermal power), the fluid pathways are reoriented to pass near the upper and lower solid regions, increasing heat transfer. As before, JAX provides seamless automatic differentiation through the coupled solvers, eliminating manual sensitivity analysis. This CHT formulation paves the way for optimizing more complex systems, such as heat exchangers, heat sinks for electronics cooling, and microfluidic devices.

\section{Conclusion}
\label{sec:conclusion}

In this work, we presented a topology optimization framework for fluidic problems that leverages automatic differentiation (AD). Implemented in Python using the JAX library, our approach simplifies the complex sensitivity analysis required for nonlinear and multiphysics systems. We demonstrated the versatility of our framework through several examples, including design of fluidic devices for minimizing dissipated power, computing optimal aerodynamic designs, the design of systems with non-Newtonian fluids, and multiphysics problems involving fluid-structure and thermo-fluid interactions. Furthermore, we highlighted how AD facilitates the seamless integration of machine learning methods for tasks such as neural network design representation and surrogate rheological modeling.

We also identify several avenues for further exposition. While the current implementation focuses on 2D, steady-state Navier-Stokes flow for simplicity, extending the framework to 3D and transient flow is of significant interest.  In addition, while we showcased speed-ups through JIT compilation and vectorization, we aim to explore more advanced performance enhancements using data-based methods or physics-informed iterative solvers. Finally, we hope the provided open-source codebase will serve as a springboard for researchers to further advance the topology optimization of complex fluidic systems.

\section*{Appendix: Implicit Automatic Differentiation}
\label{sec:appendix_implicitAutoDiff}

The nonlinear residual equations of the form in $\cref{eq:momentum}$ and $\cref{eq:continuity}$ lack explicit relationship between the parameter $\boldsymbol{\gamma}$ and the state variable $\boldsymbol{u}$. To solve such systems, we employ iterative solvers such as the Newton-Raphson (NR) method. The NR method linearizes the residual at each iteration and updates the state variable until a convergence criterion is met \cite{wengert1964simple, deledalle2014stein, domke2012generic}. The process starts with an initial guess $\boldsymbol{u^{(0)}}$ and iteratively applies the update:
\begin{equation}
\boldsymbol{u}^{(k+1)} = \boldsymbol{u}^{(k)} - \Delta \boldsymbol{u}^{(k)}
\label{eq:newton_raphson}
\end{equation}
where the update step $\Delta \boldsymbol{u}^{(k)}$ is calculated using the inverse of the Jacobian $\boldsymbol{J}$:

\begin{equation}
\Delta \boldsymbol{u}^{(k)} = J(\boldsymbol{\gamma}, \boldsymbol{u}^{(k)})^{-1} R(\boldsymbol{\gamma}, \boldsymbol{u}^{(k)}) \quad \text{where} \quad J(\boldsymbol{\gamma}, \boldsymbol{u}^{(k)}) = \frac{\partial \boldsymbol{R}}{\partial \boldsymbol{u}^{(k)}}.
\label{eq:j_inv_and_def}
\end{equation}
Iterations continue until the residual $R(\boldsymbol{\gamma}, \boldsymbol{u}^{(k)})$ is below a tolerance or a maximum number of iterations is reached.

For gradient-based optimization, we need to perform a sensitivity analysis after finding the converged solution $\boldsymbol{u}^{(K)}$. This requires the total derivative of an objective $c$ with respect to the parameter $\boldsymbol{\gamma}$:

\begin{equation}
\frac{\mathrm{d} c}{\mathrm{d} \boldsymbol{\gamma}} = \frac{\partial c}{\partial \boldsymbol{\gamma}} + \frac{\partial c}{\partial \boldsymbol{u}^{(K)}} \frac{\mathrm{d} \boldsymbol{u}^{(K)}}{\mathrm{d} \boldsymbol{\gamma}}.
\end{equation}

While the partial derivatives are straightforward, computing the parameter Jacobian, $\frac{\mathrm{d} \boldsymbol{u}^{(K)}}{\mathrm{d} \boldsymbol{\gamma}}$, is challenging. A naive approach would unroll the solver's iterations and apply the chain rule, a computationally expensive and memory-intensive process.

To avoid this, we use the Implicit Function Theorem (IFT). The IFT provides a direct way to compute the derivative from the final converged solution. Given that $R(\boldsymbol{\gamma}, \boldsymbol{u}^{(K)}) = \boldsymbol{0}$ and the Jacobian $\frac{\partial \boldsymbol{R}}{\partial \boldsymbol{u}}$ is invertible, the derivative is \cite{blondel2022efficient}:

\begin{equation}
\frac{\mathrm{d} \boldsymbol{u}^{(K)}}{\mathrm{d} \boldsymbol{\gamma}} = -\left(\frac{\partial R}{\partial \boldsymbol{u}}(\boldsymbol{\gamma}, \boldsymbol{u}^{(K)})\right)^{-1} \frac{\partial R}{\partial \boldsymbol{\gamma}}(\boldsymbol{\gamma}, \boldsymbol{u}^{(K)}).
\label{eq:IFT_derivative}
\end{equation}

In practice, particularly in frameworks such as JAX, we avoid forming the full Jacobian matrix. Instead, we compute its product with a vector $\boldsymbol{v}$ using Jacobian-vector products (JVPs) for efficiency \cite{blondel2022efficient}. The JVP formulation is:

\begin{equation}
\frac{\mathrm{d} \boldsymbol{u}^{(K)}}{\mathrm{d} \boldsymbol{\gamma}} \cdot \boldsymbol{v} = -\left(\frac{\partial \boldsymbol{R}}{\partial \boldsymbol{u}} (\boldsymbol{\gamma}, \boldsymbol{u}^{(K)}) \right)^{-1} \left(\frac{\partial \boldsymbol{R} (\boldsymbol{\gamma}, \boldsymbol{u}^{(K)})}{\partial \boldsymbol{\gamma}} \cdot \boldsymbol{v}\right).
\label{eq:jvp_formulation}
\end{equation}

\section*{Acknowledgments}
This work was supported by the Department of Mechanical Engineering at the University of Wisconsin-Madison. A.C. conducted this work as a graduate student at the University of Wisconsin-Madison.

\section*{Compliance with ethical standards}
On behalf of all authors, the corresponding author states that there is no conflict of interest.

\section*{Replication of Results}
The Python code is available at \href{https://github.com/UW-ERSL/TOFLUX}{github.com/UW-ERSL/TOFLUX}

\bibliographystyle{unsrt}  
\bibliography{references}

\end{document}